\documentstyle[aps,preprint,epsf,fixes]{revtex}

\def\d{\partial}
\def\P{{\bf P}}
\def\k{{\bf k}}

\def\q{{\bf q}}

\def\x{{\bf x}}
\def\y{{\bf y}}
\def\<{\langle}
\def\>{\rangle}

\def\max{{\text{max}}}
\def\psibar{{\overline{\psi}}}
\def\Psibar{{\overline{\Psi}}}
\def\qpar{q_\parallel}
\def\qperp{q_\perp}
\def\pf{{p_{\text{F}}}}
\def\LQCD{{\Lambda_{\text{QCD}}}}
\def\Nc{{N_{\text{c}}}}
\def\Nf{{N_{\text{f}}}}
\def\sL{{s_{\text{L}}}}
\def\EL{E_{\text{L}}}
\def\mucrit{{\mu_{\text{crit}}}}

\input prepictex
\input pictex
\input postpictex
\newdimen\tdim
\tdim=\unitlength
\def\stpltsmbl{\setplotsymbol ({\small .})}
\def\tarrow{\arrow <5\tdim> [.3,.6]}
\newbox\phdl
\setbox\phdl=\hbox{\beginpicture
\setcoordinatesystem units <\tdim,\tdim>
\stpltsmbl
\setquadratic
\plot
0 0
-3 -2.5
0 -5 
3 -7.5 
0 -10 
/
\endpicture}
\def\photondl #1 #2 *#3 /{\multiput {\copy\phdl}  at
#1 #2 *#3 0 -10 /}

\begin{document}
\setcounter{page}{0}
\def\footnoterule{\kern-3pt \hrule width\hsize \kern3pt}
\tighten

\title{On finite-density QCD at large $\Nc$}

\author{E.~Shuster\footnote{Email address: {\tt eugeneus@mit.edu}} and
D.~T.~Son\footnote{Email address: {\tt son@ctp.mit.edu}}}

\address{Center for Theoretical Physics \\
Laboratory for Nuclear Science \\
and Department of Physics \\
Massachusetts Institute of Technology \\
Cambridge, Massachusetts 02139 \\
{~}}

\date{MIT-CTP-2865,~  hep-ph/9905448. {~~~~~} May 1999}
\maketitle

\thispagestyle{empty}

\begin{abstract}

Deryagin, Grigoriev, and Rubakov (DGR) have shown that in
finite-den\-si\-ty QCD at infinite $\Nc$ the Fermi surface is unstable
with respect to the formation of chiral waves with wavenumber twice the
Fermi momentum, while the BCS instability is suppressed.  We show here
that at large, but finite $\Nc$, the DGR instability only occurs in a
finite window of chemical potentials from above $\LQCD$ to
$\mucrit\sim\exp(\gamma\ln^2\Nc + O(\ln\Nc\ln\ln\Nc))\LQCD$, where
$\gamma\approx0.02173$.  Our analysis shows that, at least in the
perturbative regime, the instability occurs only at extremely large $\Nc$,
$\Nc\gtrsim1000\Nf$, where $\Nf$ is the number of flavors.  We conclude
that the DGR instability is not likely to occur in QCD with three colors,
where the ground state at finite density is expected to be a color
superconductor.  We speculate on the possible structure of the ground
state of finite-density QCD with very large $\Nc$. 

\end{abstract}

\vspace*{\fill}
\pacs{xxxxxx}

\section{Introduction}
\label{sec:intro}

In contrast with finite-temperature QCD, QCD at high baryonic densities
remains remarkably poorly understood.  One of the main reasons is the lack
of lattice simulations due to the complex fermion determinant in
finite-density QCD. Meanwhile, the physics in the core of the neutron
stars, and possibly of heavy-ion collisions, depends crucially on the
structure and properties of the ground state of QCD at finite densities. 

It was suggested that at sufficiently high densities, the ground state of
QCD is a color superconductor \cite{BailinLove,super}.  Such state arises
from the instability of the Fermi surface under the formation of Cooper
pairs of quarks.  The superconducting phase of quark matter is the subject
of many recent studies \cite{super2} and we will not discuss its
properties in this paper.  We will only note that a reliable treatment is
currently available only in the perturbative regime of asymptotically high
densities \cite{Son}; in the physically most interesting regime of
moderate densities, QCD is strongly coupled and one has to resort to
various toy models, e.g.\ those with four-fermion interactions
\cite{super}. 

To shed light on possible new phases that may occur in the
non-perturbative regime of moderate baryonic densities, one might hope to
be able to make use of alternative limits, such as the large $\Nc$ limit,
where one takes the number of colors $\Nc$ to infinity, keeping $g^2\Nc$
fixed ($g$ is the gauge coupling) \cite{Witten}.  This limit has proved to
be a convenient framework for understanding many properties of QCD (for
example, Zweig rule), although QCD at infinite $\Nc$ is still not
analytically treatable. In the context of finite-density QCD, the first
work that discussed the implications of the large $\Nc$ limit was done by
Deryagin, Grigoriev, and Rubakov (DGR) \cite{DGR}.  DGR noticed that color
superconductivity is suppressed at large $\Nc$ due to the fact that the
Cooper pair is not a color singlet (the diagram responsible for color
superconductivity is non-planar).\footnote{At arbitrary $\Nc$, using the
technique of Ref.\ \cite{Son},the asymptotic behavior of the BCS gap can
be found to be $\Delta\sim\mu\exp\biggl(-\sqrt{6\Nc\over\Nc+1}{\pi^2\over
g}\biggr)$.  This tends to 0 as $\Nc\to\infty$, provided one keeps
$g^2\Nc$ fixed.} Working in the perturbative regime $g^2\Nc\ll1$, DGR
noticed another instability of the Fermi surface, this time with respect
to the formation of chiral waves with wavenumber $2\pf$, where $\pf$ is
the Fermi momentum.  As shown by DGR, this instability is not suppressed
in the limit $\Nc\to\infty$. 

The purpose of this paper is to see what happens to the DGR instability at
large but finite $\Nc$.  Our motivation is to see whether the limit
$\Nc\to\infty$ is relevant for the physics of high-density QCD at $\Nc=3$.
In this paper, we find that at any fixed value of the chemical potential
$\mu$, in order for the DGR instability to occur we require the number of
colors $\Nc$ to be larger than some minimum value $\Nc(\mu)$, which grows
with $\mu$.  What is surprising is that even for moderate values of $\mu$,
the minimum value $\Nc(\mu)$ is very large (of order of a few thousands
for a modest chemical potential $\mu=3\LQCD$).  Therefore one should not
expect the large $\Nc$ limit to be of direct relevance for physics with
$\Nc=3$ at finite densities. 

The paper is organized as follows.  Section \ref{sec:review} reviews the
results of DGR.  A convenient technical approach to DGR instability which
is based on renormalization group is developed in Sec.\ \ref{sec:rg} and
applied to the case of finite $\Nc$ in Sec.\ \ref{sec:finitenc}.  Section
\ref{sec:concl} contains concluding remarks.

\section{Review of DGR results}
\label{sec:review}

Let us review the key results of Ref.\ \cite{DGR}.  Throughout our
paper, we assume all quarks are massless, and make no distinction between
Fermi momentum and Fermi energy: $\pf=\mu$.  In the $\Nc\to\infty$
limit, the DGR result states that the Fermi surface is unstable under
the development of chiral waves with wavenumber $2\mu$,
\begin{equation}
  \< \psibar(x)\psi(y) \> = e^{i\P\cdot(\x+\y)}
  \int\!d^4q\, e^{-iq(x-y)} f(q)
  \label{DGR}
\end{equation}
where $\P$ is a vector with modulus $|\P|=\mu$ whose direction is fixed
arbitrarily.  Since $\psibar\psi$ is a color singlet, it survives the
limit $\Nc\to\infty$.
\begin{figure} \centering
$$
      \def\epsfsize #1#2{0.7#1}
      \epsfbox{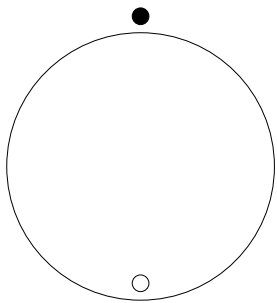}
$$
\caption{The particle-hole pair} \label{fig:pair}
\end{figure}

The condensate (\ref{DGR}) can be interpreted as the formation of
particle-hole pairs with total momentum $2\P$ (Fig.\ \ref{fig:pair}).  In
such a pair, both the particle and the hole are near the Fermi surface,
and the momenta of the particle and the hole are both near $\P$.  In this
sense the condensate (\ref{DGR}) is different from the usual chiral
condensate $\<\psibar\psi\>=\text{const}$, which corresponds to the
pairing of a particle and a antiparticle moving in opposite directions. 
Moving in the same directions, the scattering between the particle and the
hole is nearly in the foward direction, and since the amplitude of forward
scattering is singular, one could expect the formation of the pair to be
energetically favorable.  In fact, this is the reason why the total
momentum $2\mu$ is special. 

The function $f(q)$ has the physical meaning of the wave function of the
pair in the center-of-mass frame, so $\P+\q$ is the momentum of the
particle and $\P-\q$ is that of the hole.  DGR found that the wave
function is localized in an exponentially small region of momenta
$q<\Delta_\perp$ where
\begin{equation}
  \Delta_\perp \simeq \mu e^{-\pi/2h},\qquad h^2={g^2\Nc\over4\pi^2}.
  \label{DGR_Delta}
\end{equation}
Recall that $h$ is kept constant in the limit $\Nc\to\infty$.  The binding
energy of the pair is found to be at an even smaller scale,
\begin{equation}
  E_{\text{bind}} \sim \mu e^{-\pi/h}.
  \label{DGR_binding}
\end{equation}
Both scales $\Delta_\perp$ and $E_{\text{bind}}$ are parametrically
larger than the non-perturbative scale $\LQCD\sim\mu e^{-6/11h^2}$. For
more details, see Ref.\ \cite{DGR}. 

It may seem surprising that the DGR instability occurs in the perturbative
regime.  Indeed, the analogs of (\ref{DGR}), in non-relativistic fermion
systems, are the charge-density wave (CDW) and the spin-density wave
(SDW).  Since it is known that in three dimensions CDW and SDW do not
develop at small four-fermion interaction, one could ask how such
instability could occur at small $g^2\Nc$.  The key observation is that in
our case the effective four-fermion interaction is singular due to the
$1/q^2$ behavior of the gluon propagator at small $q$. In general, this
singularity is cut off by screening, but because the diagrams
responsible for screening involve fermion loops, the screening effects
at large $\Nc$ are of order $g^2 \mu^2 \sim O(1/\Nc)$ and therefore 
suppressed. This singular nature of the interaction explains why the DGR
instability can occur perturbatively at large $\Nc$.

The argument presented above also implies that at each value of the
coupling $h$, there must be a lower limit on $\Nc$, below which the
interaction is not singular enough due to the screening, and the DGR
instability disappears.  This limit grows as one decreases $h$, or,
equivalently, as the chemical potential increases.  Finding this lower
bound on $\Nc$ as a function of $\mu$ is the purpose of this paper.

\section{Renormalization group approach to DGR instability}
\label{sec:rg}

Before tackling our main problem, let us formulate an efficient RG
technique that reproduces the results of DGR in the $\Nc\to\infty$
limit. While in the limit $\Nc\to\infty$ this technique does not give us
anything new over what has been already found by DGR, it has the
advantage that it can be applied to the case of finite $\Nc$, where the
effects of screening make the generalization of the original method of
Ref.\ \cite{DGR} very difficult, if at all possible.  We will not try to
rigorously justify the RG in this paper.

\begin{figure}
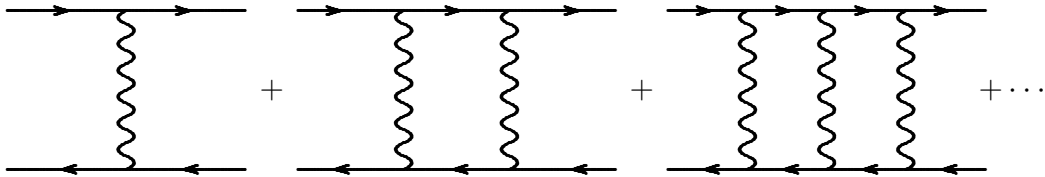
$$
\beginpicture
\setcoordinatesystem units <\tdim,\tdim>
\stpltsmbl
\plot -200 30 -110 30 /
\plot -200 -30 -110 -30 /
\photondl -155 30 *5 /
\tarrow from -179 30 to -177 30
\tarrow from -131 -30 to -133 -30
\tarrow from -133 30 to -131 30
\tarrow from -177 -30 to -179 -30

\plot -90 30 30 30 /
\plot -90 -30 30 -30 /
\photondl -50 30 *5 /
\photondl -10 30 *5 /
\tarrow from -76 30 to -74 30
\tarrow from 16 -30 to 14 -30
\tarrow from -31 30 to -29 30
\tarrow from -29 -30 to -31 -30
\tarrow from 14 30 to 16 30
\tarrow from -74 -30 to -76 -30

\plot 50 30 170 30 /
\plot 50 -30 170 -30 /
\photondl 80 30 *5 /
\photondl 110 30 *5 /
\photondl 140 30 *5 /
\tarrow from 64 30 to 66 30
\tarrow from 96 -30 to 94 -30
\tarrow from 124 30 to 126 30
\tarrow from 156 -30 to 154 -30
\tarrow from 154 30 to 156 30
\tarrow from 126 -30 to 124 -30
\tarrow from 94 30 to 96 30
\tarrow from 66 -30 to 64 -30

\put {$+$} at -100 0
\put {$+$} at 40 0
\put {$+\cdots$} at 180 0

\endpicture$$
\vspace*{5pt}
\caption{Ladder approximation}
\label{fig:ladder}
\end{figure}

Let us stay in the Fermi liquid phase, where quarks are deconfined, and
consider the scattering between a particle and a hole with momenta $\P+\q$
and $\P-\q$.  The total momentum of the pair is $2\mu$.  A singularity of
this scattering amplitude in the upper half of the complex energy plane
would signify an exponentially growing mode, i.e.  an instability
\cite{AGD}.  In terms of the diagrams, the most important contribution to
the scattering amplitude comes from the ladder graphs (Fig.\
\ref{fig:ladder}).  Adding a rung to the ladder brings two more
logarithms: one comes from the collinear divergence, i.e.\ the singular
gluon propagator, and the other from the fact that the two new fermion
propagators are near the mass shell.  We will design the RG to resum these
double logs.\footnote{A similar but not identical RG procedure has been
developed to resum the double logs in the BCS channel \cite{Son}.}

Our first step is to derive a 1+1 dimensional effective theory capable of
describing the DGR instability.  On the most naive level, such description
exists due to the fact that the modes of interest move in directions close
to $\pm\P$.  Technically, the (1+1)D effective theory arises from
integrating, in each Feynman graph, over the momentum components
perpendicular to $\P$ \cite{Altshuler}.

\begin{figure}
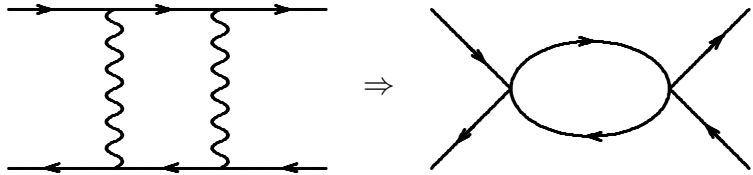
$$
\beginpicture
\setcoordinatesystem units <\tdim,\tdim>
\stpltsmbl
\plot -190 30 -70 30 /
\plot -190 -30 -70 -30 /
\photondl -150 30 *5 /
\photondl -110 30 *5 /
\tarrow from -176 30 to -174 30
\tarrow from -84 -30 to -86 -30  
\tarrow from -131 30 to -129 30
\tarrow from -129 -30 to -131 -30
\tarrow from -86 30 to -84 30
\tarrow from -174 -30 to -176 -30

\ellipticalarc axes ratio 5:3 360 degrees from 60 0 center at 30 0
\plot -30 -30 0 0 -30 30 /
\plot 90 -30 60 0 90 30 /
\tarrow from -14 -14 to -20 -20
\tarrow from -16 16 to -10 10
\tarrow from 74 14 to 80 20
\tarrow from 80 -20 to 74 -14
\tarrow from 29 18 to 31 18
\tarrow from 31 -18 to 29 -18

\put {$\Rightarrow$} at -50 0

\endpicture$$
\vspace*{5pt}
\caption{Reducing one-gluon exchange to a point-like interaction in the
effective theory}
\label{fig:reduction}
\end{figure}

Let us consider a ladder graph and ask what happens if one adds one more
rung.  The diagram now contains an extra loop integral,
\begin{equation}
  \int\!{d^4q\over(2\pi)^4}\,G(P+q)G(-P+q)D(q)  
  \label{loop-int}
\end{equation}
where $G$ and $D$ are fermion and gluon propagators respectively, and
the Dirac structure of the fermion propagators is ignored for the
purposes of the discussion presented below.  Consider first the  
fermion line with momentum $\P+\q$. The component of the fermion momentum
parallel to $\P$ will be denoted as $\mu+\qpar$, and those perpendicular
to $\P$ will be denoted as $\qperp$.  Note that $\qperp$ is a
two-dimensional vector.  We will assume that for all fermion lines in the
Feynman diagram $\qperp\sim\Delta$, where $\Delta$ is an arbitrary
momentum scale much less than $\mu$.  In other words, we will be
interested only in the modes located inside two small ``patches'' on the
Fermi sphere, each having the size of order $\Delta$ in directions
perpendicular to $\P$ (eventually, $\Delta$ will be identified with
$\Delta_\perp$ in Eq.\ (\ref{DGR_Delta})).  When $\qpar$ is also small
compared to $\mu$, the fermion propagator has the form
\begin{equation}
  G(q) \sim {1\over iq_0+|\P+\q|-\mu} \approx
  {1\over iq_0+\qpar+{\qperp^2\over2\mu}}\, .
  \label{parprop}
\end{equation}
If $\qpar\gg\qperp^2/\mu\sim\Delta^2/\mu$, the $\qperp$ dependence drops
out and the propagator is simply $(iq_0+\qpar)^{-1}$.  Therefore, in the
regime $\qpar\gg\Delta^2/\mu$, the fermion propagator does not depend on
the perpendicular (with respect to $\P$) momenta.  In this regime, in Eq.\
(\ref{loop-int}) only the gluon propagator $D(q)$ depends on
$\qperp$. Hence, the integration over $\qperp$ has the form
\begin{equation}
  \int\!
  {d^2\qperp\over(2\pi)^2}\, {1\over q_0^2+\qpar^2+\qperp^2}\, .
  \label{intparperp}
\end{equation}
If $q_0$ and $\qpar$ are not only small compared to $\mu$, but also much
smaller than $\Delta$, then the integral over $\qperp$ in Eq.\
(\ref{intparperp}) is a logarithmic one $\int\!d^2\qperp/\qperp^2$.
The integral is cut off in the IR by $\qpar$ and in the 
UV by $\Delta$ and yields ${g^2\over4\pi}\ln{\Delta\over\qpar}$.
Effectively, this integration replaces the internal gluon line by a
four-fermion vertex ${g^2\over4\pi}\ln{\Delta\over\qpar}$, where $\qpar$
is determined by the momentum of the fermions coming in and out of the
vertex (Fig.\ \ref{fig:reduction}). Recall that the simplification
takes place only in the region $\Delta\gg\qpar\gg\Delta^2/\mu$, as
only in this region the integration over $\qperp$ decouples from that
over  $\qpar$. At the end of this section we argue why the restriction
of $\qpar$ to the region $\Delta^2/\mu \ll \qpar \ll \Delta$ is well
justified.

We have taken the integration over the perpendicular components of one
particular gluon momentum, but nothing prevents us from integrating over
the perpendicular components of {\em all} the gluon momenta.  By doing this
integration, we resum one set of logarithms (the one related to the
collinear divergence) in a series of double logs.  Now, as the only
remaining integrals are over $q_0$ and $\qpar$, all Feynman diagrams
are identical to those of some 1+1 dimensional model with a four-fermion
interaction.  Our task is to find out the precise form of the Lagrangian
of this model.

First, we note that the kinetic term for the fermions in the effective
theory can be obtained from the original Lagrangian by omitting spatial
derivatives in directions other than $z$,
\begin{equation}
  L_{\text{kin}} = i\psibar\gamma^0\d_0\psi + i\psibar\gamma^3\d_3\psi + 
      \mu\psibar\gamma^0\psi.
  \label{2dmid} 
\end{equation}
It is more convenient, however, to recast the Lagrangian (\ref{2dmid}) 
into the form of a (1+1)D theory of a doublet of Dirac fermions (which are
two-component in (1+1)D)  at {\em zero} chemical potential.  This is
indeed possible, since spinless fermions at finite chemical potential can
be rewritten as one Dirac fermion at zero chemical potential (the modes
near two points of the ``Fermi surface'' serve as its two components
\cite{Tsvelik}).  It is not surprising that in our case the
spin-${1\over2}$ fermions can be rewritten as a doublet of (1+1)D Dirac
fermions.  Let us do it explicitly when $\P$ is directed along the
$z$-axis, $\P=(0,0,\mu)$.  Denote the four components of the Dirac spinor
$\psi$ (in chiral basis)  as $\psi^{\rm T} = (\psi_{\rm L1}, \psi_{\rm
L2}, \psi_{\rm R1}, \psi_{\rm R2})$. The antiparticles have energy of
order $2\mu$ and decouple from the low-energy effective theory that is
being derived.  This allows us to consider only the components of $\psi$
corresponding to particles, which are $\psi_{\rm L2}$ and $\psi_{\rm R1}$
when the particle's momentum is near $\P$, and $\psi_{\rm L1}$ and
$\psi_{\rm R2}$ when it is near $-\P$.  Although these fields are slowly
varying in time, they still vary rapidly in space. To compensate for this
spatial variation, we introduce new fields,
\begin{equation}
  \varphi = \left( \begin{array}{c}
     e^{-i\mu z} \psi_{\rm L2} \\ e^{i\mu z}\psi_{\rm R2}
  \end{array}\right), \qquad
  \chi = \left( \begin{array}{c}
     e^{-i\mu z} \psi_{\rm R1} \\ e^{i\mu z}\psi_{\rm L1}
  \end{array}\right)
  \label{phichi}
\end{equation}
which are soft in both space and time.  We can now translate from the
(3+1)D language of $\psi$ to the (1+1)D language of $\varphi$ and $\chi$. 
The kinetic part of the Lagrangian (\ref{2dmid}) becomes
\begin{equation}
  L_{\text{kin}} = i\psibar\gamma^0\d_0\psi + i\psibar\gamma^3\d_3\psi + 
      \mu\psibar\gamma^0\psi\to
  i\overline{\varphi}\gamma^\mu_{\text{2D}}\d_\mu\varphi +
  i\overline{\chi}\gamma^\mu_{\text{2D}}\d_\mu\chi.
  \label{trans_kin}
\end{equation}
What is the interaction term in the effective theory?  A look at the
Feynman diagram in Fig.\ \ref{fig:reduction} tells us that such
interaction is of the current-current type.  The current operator can also
be translated into the (1+1)D counterparts,
\begin{equation}
  \psibar\gamma^\mu\psi \to
  \overline{\varphi}\gamma^\mu_{\text{2D}}\varphi+
  \overline{\chi}\gamma^\mu_{\text{2D}}\chi
  \label{trans_curr}
\end{equation}
where $\gamma^\mu_{\text{2D}}$ are two (1+1)D Dirac matrices,
$\gamma^0_{\text{2D}}=\sigma^1$, $\gamma^1_{\text{2D}}=-i\sigma^2$.  Below
we will write these matrices simply as $\gamma^\mu$ in all expressions
belonging to the (1+1)D effective theory.  Noting that each vertex in
Fig.\ \ref{fig:reduction} corresponds to a factor of
${g^2\over4\pi}\ln{\Delta\over\qpar}$, where $\qpar$ is the parallel
momentum transfer, we find that the Lagrangian of the (1+1)D effective
theory is similar to that of the non-Abelian Thirring model
\begin{equation}
  L_{\text{eff}} = i\Psibar\gamma^\mu\d_\mu\Psi-
  {g^2\over4\pi}\ln{\Delta\over\qpar}
  \biggl(\Psibar\gamma^\mu{T^a\over2}\Psi\biggr)^2.
  \label{eff_theory}
\end{equation}
where we have combined the two fields $\varphi$ and $\chi$ into a doublet
$\Psi$.  The only difference between (\ref{eff_theory}) and the
non-Abelian Thirring model is the dependence of the four-fermion coupling
on the scale of the parallel momentum exchange $\qpar$.  The theory
(\ref{eff_theory})  describes the interaction between fermions with
perpendicular momenta of order $\Delta$ and parallel momenta between
$\Delta^2/\mu$ and $\Delta$. 

\begin{figure}
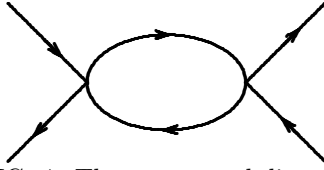
$$
\beginpicture
\setcoordinatesystem units <\tdim,\tdim>
\stpltsmbl
\ellipticalarc axes ratio 5:3 360 degrees from 30 0 center at 0 0
\plot -60 -30 -30 0 -60 30 /
\plot 60 -30 30 0 60 30 /
\tarrow from -44 -14 to -50 -20
\tarrow from -46 16 to -40 10
\tarrow from 44 14 to 50 20
\tarrow from 50 -20 to 44 -14
\tarrow from -1 18 to 1 18
\tarrow from 1 -18 to -1 -18
\endpicture$$
\vspace*{5pt}
\caption{The zero-sound diagram}
\label{fig:zerosound}
\end{figure}

To understand the properties of the model (\ref{eff_theory}), let us
recall what is known about the conventional Thirring model, where the
interaction term is $-\lambda(\Psibar\gamma^\mu{T^a\over2}\Psi)^2$.
The Thirring model is asymptotically free.  The only diagram contributing
to the $\beta$ function at large $\Nc$ is the ``zero-sound'' diagram,
Fig.\ \ref{fig:zerosound}.
The running of the coupling $\lambda$ is governed by the RG equation,
\[
  {\d\lambda(s)\over\d s}= {\Nc\over\pi}\lambda^2(s)
\]
where $s$ is the RG parameter, and $\lambda(s)$ is the coupling at the
energy scale $\Delta e^{-s}$.  The coupling $\lambda$ hits a Landau pole
at $p\sim\Delta e^{-\pi/\lambda\Nc}$.  The physics in the IR is
characterized by the formation of the chiral condensate $\<\Psibar\Psi\>$
which gives mass to the fermions.  Using Eq.\ (\ref{phichi}), one can see
that $\<\Psibar\Psi\>=\cos2\mu z\<\psibar\psi\>-i\sin2\mu
z\<\psibar\gamma^0\gamma^3\psi\>$, so a constant $\<\Psibar\Psi\>$
translates into space-dependent condensates $\<\psibar\psi\>$ and
$\<\psibar\gamma^0\gamma^3\psi\>$.

These basic properties hold for the model (\ref{eff_theory}) as well, but
the estimation for the scale of the Landau pole is different. The latter
can be found using RG.  Now the RG equation needs to be written for a
coupling which is a function of the parallel momentum transfer $\qpar$. At
$s=0$,
\begin{equation}
  \lambda(\qpar) = {g^2\over4\pi}\ln{\Delta\over\qpar} \, .
  \label{RGinitp}
\end{equation}
The RG equation is found from the diagram drawn in Fig.\
\ref{fig:zerosound}.  The internal fermion lines have the momentum of
order $\Delta e^{-s}$, which is much larger than the momentum of the
external lines, therefore the momentum transfer at each vertex is
$\Delta e^{-s}$.  The RG equation, therefore, is
\[
  {\d\over\d s}\lambda(s,\qpar) = {\Nc\over\pi}\lambda^2(s,\Delta e^{-s}).
\]
It is convenient to use the logarithmic parameter $u$, defined by
$\qpar = \Delta e^{-u}$, and rewrite the RG equation as
\begin{equation}
  {\d\over\d s} \lambda(s,u) = {\Nc\over\pi}\lambda^2(s,s).
  \label{RGl}
\end{equation}
The initial condition (\ref{RGinitp}) becomes
\begin{equation}
  \lambda(0,u) = {g^2\over4\pi}u.
  \label{RGlinit}
\end{equation}
One should note that at the moment $s$ of the RG evolution, all fermion
modes with energy larger than $\Delta e^{-s}$ have been integrated out,
therefore the function $\lambda(s,u)$ is defined only for $u>s$.  The
solution to Eq.\ (\ref{RGl}) with the initial condition Eq.\
(\ref{RGlinit}) is
\begin{equation}
  \lambda(s,u) = {\pi\over\Nc}f(s) + {g^2\over4\pi}(u-s)
  \label{lambdaf}
\end{equation}
where $f(s)$ satisfies the equation
\begin{equation}
  {\d\over\d s} f(s) = h^2 + f^2(s)
  \label{df/dt}
\end{equation}
and $h^2 = g^2\Nc/4\pi^{2}$.  Solving Eq.\ (\ref{df/dt}) one finds $f(s) =
h\tan hs$, which hits a Landau pole at $s=\sL=\pi/2h$.  The corresponding
scale is $\EL=\Delta e^{-\pi/2h}$. Recall now that RG evolution occurs for
$\Delta^2/\mu \ll \qpar \ll \Delta$. From this condition one finds that
the Landau pole can only be achieved if $\Delta^2/\mu \lesssim\EL$ or
$\Delta\lesssim\mu e^{-\pi/2h}$.  Under this constraint, the maximal value
of $\EL$ is achieved when $\Delta\sim\mu e^{-\pi/2h}$, at which $\EL =
\Delta^2/\mu \sim \mu e^{-\pi/h}$.  Thus, the estimation for the Landau
pole scale $\EL$ and for $\Delta$ coincide with the result found by DGR
for the binding energy of the particle-hole pair and the size of the pair
wave function, Eqs.\ (\ref{DGR_Delta},\ref{DGR_binding}).

Now, it is easy to demonstrate why we were justified to consider only the
region $\Delta^2/\mu \ll \qpar \ll \Delta$ in the argument presented
above. On one hand, when $\qpar$ drops below the scale $\Delta^2/\mu$,
we cannot neglect the dependence of the fermion propagator on $\qperp$,
which now acts as a cut off for the RG flow. Hence, for $\qpar \lesssim
\Delta^2/\mu$, there is no RG flow in the effective (1+1)D theory and
the Landau pole is never reached. On the other hand, when $\qpar$
becomes comparable with $\qperp$ (i.e. $\Delta$), we cannot neglect
$\qpar$ dependence in the gluon propagator.  One can estimate the effect
of such dependence by noticing that the four-fermion coupling in the
effective (1+1)D theory (\ref{eff_theory}) now reads
\[
  \lambda(\qpar) =
  {g^2\over8\pi}\ln\left(1+{\Delta^2\over\qpar^2}\right)
\]
and the RG equation (\ref{df/dt}) becomes
\begin{equation}
  {\d\over\d s} f(s) = {h^2\over{1+e^{-2s}}} + f^2(s).
\end{equation}
One can see that for $\qpar \gtrsim \Delta$, the RG flow in the effective
(1+1)D is completely negligible.  Therefore, to find the DGR instability
we can restrict the values of $\qpar$ to lie between $\Delta^2/\mu$ and
$\Delta$. 

Having reproduced the DGR results by our RG procedure, let us turn to the
case of large, but finite $\Nc$.

\section{DGR instability at finite $\Nc$}
\label{sec:finitenc}

The RG technique described above can be very easily extended to the case
of large but finite $\Nc$.  The effect of finite $\Nc$ is to cut off the
IR singularity of the gluon propagator at small momentum exchange by
Thomas-Fermi screening and Landau damping.  The electric propagator
becomes $(q^2+m^2)^{-1}$, and the magnetic propagator becomes
$(q^2+im^2|q_0|/q)^{-1}$ \cite{LeBellac}, where $m$ is the Thomas-Fermi
screening scale of order $g\mu$.  If the screening mass $m$ is smaller
than the scale of the Landau pole found in Sec.\ \ref{sec:rg}, i.e.\ $\mu
e^{-\pi/h}$, then our previous calculations are not affected. However, if
$m> \mu e^{-\pi/h}$, we need to modify the RG to take into account the
screening.

The screening affects the integration over perpendicular components of the
gluon propagators: before these integrals were cut off by the parallel
exchanged momentum $\qpar$, now it is cut off by the largest scale among
$\qpar$ and $m$ in the case of electric gluons, and among $\qpar$ and
$m^{2/3}\qpar^{1/3}$ in the case of magnetic gluons.  The effective (1+1)D
theory is now a Thirring-like model with different scale-dependent
couplings for the electric and magnetic interactions,
\begin{equation}
  L_{\text{eff}} = i\Psibar\gamma^\mu\d_\mu\Psi
  -\lambda_0(\qpar)\biggl(\Psibar\gamma^0{T^a\over2}\Psi\biggr)^2
  +\lambda_1(\qpar)\biggl(\Psibar\gamma^1{T^a\over2}\Psi\biggr)^2.
  \label{eff_theory2}
\end{equation}
where
\begin{eqnarray*}
  \lambda_0(\qpar) & = & {g^2\over4\pi}\ln{\Delta\over\max(\qpar,m)} \\
  \lambda_1(\qpar) & = &
     {g^2\over4\pi}\ln{\Delta\over\max(\qpar,m^{2/3}\qpar^{1/3})}.
\end{eqnarray*}
The RG equations for $\lambda_+=(\lambda_0+\lambda_1)/2$ and
$\lambda_-=(\lambda_0-\lambda_1)/2$ decouple:
\begin{eqnarray}
  {\d\over\d s}\lambda_+(s,u) & = & {\Nc\over\pi}\lambda_+^2(s,s) 
  \label{RGl+}\\
  {\d\over\d s}\lambda_-(s,u) & = & 0. \nonumber
\end{eqnarray}
where again $u=\ln{\Delta\over\qpar}$.  Therefore, only $\lambda_+$
changes during the RG evolution.  The initial condition for $\lambda_+$
can be read from Eq.\ (\ref{eff_theory2}),
\begin{equation}
  \lambda_+(0,u) = \left\{ \begin{array}{ll}
    \displaystyle{{g^2\over4\pi}u} & \mbox{if $u<s_m$} \\
    \displaystyle{{g^2\over4\pi} \biggl({5\over6}s_m+{1\over6}u\biggr)} & 
    \mbox{if $u>s_m$} 
  \end{array} \right.
  \label{l+init}
\end{equation}
where $s_m = \ln{\Delta\over m}$.  The solution to Eq.\ (\ref{RGl+}) with
the initial condition (\ref{l+init}) can be written in the form of Eq.\
(\ref{lambdaf}), where $f(s)$ now satisfies the equation
\[
  {\d\over\d s}f(s) = \left\{ \begin{array}{ll}
    f^2 + h^2 & \mbox{if $s<s_m$} \\
    \displaystyle{f^2 + {h^2\over6}} & \mbox{if $s>s_m$}
  \end{array} \right. \, .
\]
The solution to this equation is
\[
  f(s) = \left\{ \begin{array}{ll}
    h\tan hs & \mbox{if $s<s_m$} \\
    \displaystyle{{h\over\sqrt{6}}\tan{h\over\sqrt{6}}(s+c)} & 
    \mbox{if $s>s_m$}
  \end{array} \right.
\]
where $c$ can be found by matching the solution at $s=s_m$:
\[
  c = {\sqrt{6}\over h}\arctan(\sqrt{6}\tan hs_m) - s_m.
\]
The Landau pole occurs at
\[
  \sL = {\sqrt{6}\pi\over2h}-c =
  {\sqrt{6}\over h}\arctan({1\over\sqrt{6}}\cot hs_m) + s_m.
\]
Recall that for the instability to really occur, the scale of the Landau
pole should be larger than the scale $\Delta^2/\mu$, one finds a condition
on $m$,
\[
  m = \Delta e^{-s_m} < \mu e^{-\sL-s_m} =
  \mu\exp\biggl[-{\sqrt{6}\over h}\arctan\biggl({1\over\sqrt{6}}
  \cot hs_m\biggr)-2s_m\biggr].
\]
One can maximize the right hand side (RHS) of this equation to find the
maximum value of $m$ where the Landau pole still can be achieved.  One
finds that for the Landau pole to be reached, $m$ should be smaller than
$m_\max = \mu e^{-c/h}$, where
\[
  c= \sqrt{6}\arctan{1\over2}+ 2\arctan\sqrt{2\over3} \approx 2.5051.
\]
This restriction on $m$ leads to a condition on $\Nc$ and $\mu$ for the
DGR instability to occur.  Recall that the Thomas-Fermi mass is
\[
  m = \sqrt{\Nf\over2\pi^2} g\mu
\]
(which is of order $\Nc^{-1/2}$), we see that at a fixed coupling $g^2\Nc$
(or, equivalently, $\mu$), there exists a lower bound on $\Nc$ where
condition $m<\mu e^{-c/h}$ is satisfied.  The lower bound can be easily
found to be
\begin{equation}
  \Nc \gtrsim 2\Nf h^2 e^{2c/h}.
  \label{Ncbound}
\end{equation}
Since our arguments rely on the comparison of scales, Eq.\ (\ref{Ncbound}) 
contains an extra unknown coefficient of order 1 on the RHS.  As the
chemical potential $\mu$ increases, the effective coupling $h$ decreases;
using the one-loop beta function
\begin{equation}
 h^{2} = {6\over{11\ln{\mu\over\LQCD}}}
 \label{hmudependence}
\end{equation}
and according to Eq.\ (\ref{Ncbound}) the lower bound on $\Nc$ increases. 
In reality, the numerical constant $2c$ in the exponent on the RHS of Eq.\
(\ref{Ncbound}) is relatively large ($\approx5$), so the lower bound is
already large at moderate values of $\mu$.  For example, if one uses the
value of $h$ corresponding to $\mu=3\LQCD$, the RHS of Eq.\
(\ref{Ncbound}) is of order $1000\Nf$!  Barring the possibility of a very
small numerical constant on the RHS of Eq.\ (\ref{Ncbound}), which seems
unlikely, this lower bound is always much larger than 3. 
\begin{figure} \centering
$$
      \def\epsfsize #1#2{0.7#1}
      \epsfbox{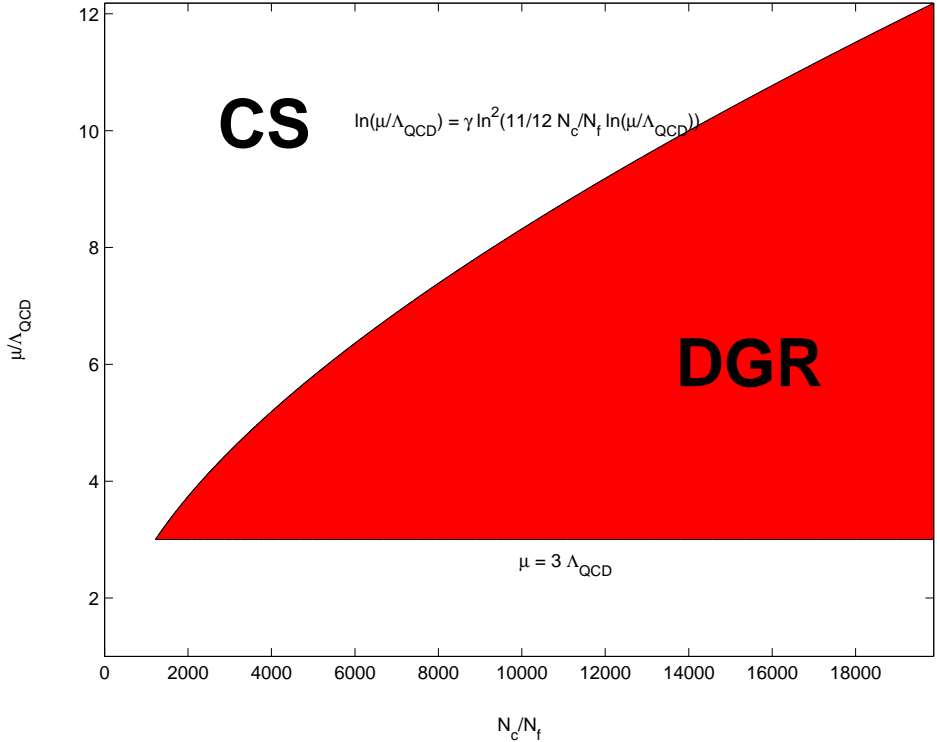}
$$
\caption{Region of DGR instability in the $(\Nc,\mu)$ plane from Eq.\
(\ref{Ncbound}).  CS and DGR stand for regions with predominant color
superconductivity and DGR instability, respectively.}
\label{fig:Ncmu}
\end{figure}
From Eqs.\ (\ref{Ncbound},\ref{hmudependence}), one can construct the
phase diagram of QCD in the $(\Nc,\mu)$ plane.  The result is shown in
Fig.\ \ref{fig:Ncmu}.  In the shaded region, $\Nc$ satisfies the
inequality (\ref{Ncbound}), which means that DGR instability occurs.  We
restrict this region by the line $\mu=3\LQCD$, for below this line QCD is
certainly strongly-coupled and not much can be said from our calculation. 
Above the curved line, inequality (\ref{Ncbound}) is not satisfied, and
the Fermi surface is stable in the DGR channel.  However, the BCS
instability is still there (though suppressed by large $\Nc$), thus
implying that the ground state of QCD is a color superconductor in that
region. 

At any given (large) $\Nc$, the DGR instability occurs only in a finite
window of the values of the chemical potential.  The maximal value of
$\mu$ where DGR instability still occurs, $\mucrit$, can be found by
solving (\ref{Ncbound}) with respect to $\mu$.  Asymptotically,
\[
  \mucrit \sim \exp(\gamma\ln^2\Nc + O(\ln\Nc\ln\ln\Nc))\LQCD \sim
  \Nc^{\gamma\ln\Nc}\LQCD
\]
where
\[
  \gamma = {3\over22c^2} = 0.02173\ldots
\]
The smallness of the numerical constant $\gamma$ and the logarithmic
dependence of $\mucrit$ on $\Nc$ are the reasons why it requires a
numerically large $\Nc$ for $\mucrit$ to be as small as
$3\LQCD$.  However, asymptotically $\mucrit$ grows faster than any power
of $\Nc$.

\section{Conclusion}
\label{sec:concl}

In this paper we have seen that in finite-density QCD the Fermi surface is
unstable under the DGR instability in a finite range of chemical
potential.  We have also found that the number of colors $\Nc$ needs to be
numerically very large for the DGR instability to occur in
perturbation theory.  This indicates that at low $\Nc$ (like $\Nc=3$), the
DGR instability might not have a chance to realize itself at any value
of the chemical potential and the only
instability of the Fermi surface is the BCS one, which leads to color
superconductivity. 

Returning to the case of very large $\Nc$, the next logical step is to
ask what is the ground state once the Fermi liquid is unstable under the
DGR particle-hole pairing.  This seems to be a purely academic exercise
due to the large $\Nc$ required, but it might still be interesting because
of the possibility, at least in principle, of a new phase, distinct from
the Fermi liquid and BCS superconducting phases in 3D fermionic systems.
In the original paper \cite{DGR}, DGR constructed a ``standing chiral wave
state'', in which $\<\psibar\psi\>$ varies periodically in space with
wavenumber $2\mu$.  This state is periodic only along one spatial
direction and does not break translational symmetry along the other two
directions. Since translational symmetry cannot be broken in only one
direction, such state cannot be the ground state of QCD.

One notices that a chiral wave with a particular wavevector utilizes only
fermion modes in a small region with size $\Delta_\perp$ (Eq.\
(\ref{DGR_Delta})) near two opposite points on the Fermi sphere.  It is
clear how to make a state with energy smaller than the original DGR
standing wave state.  Indeed, one can pair up particles and holes in
different pairs of opposite patches on the Fermi sphere.  Since the size of
each patch is exponentially small compared to the total area of the Fermi
surface, one can have a large number of patches that do not overlap with
each other.  From the size of the patches one deduces that one can place a
maximum of $e^{-\pi/h}$ patches on the sphere.  The condensate has the
form of a linear combination of $e^{i\k_i\cdot\x}$, where all $\k_i$ have
modulus equal to $2\mu$ but point in different directions.  It is easy to
estimate the energy gain from forming such a state.  Indeed, the pairing
affects fermions in a thin shell near the Fermi surface; the thickness of
the shell is the scale at which we have found the Landau pole, i.e.\ $\mu
e^{-\pi/h}$.  Therefore, the fraction of fermions affected is
$e^{-\pi/h}$, and each pair lowers the energy by $\mu e^{-\pi/h}$. 
Therefore, the gain in energy density is
\begin{equation}
  \mu^4 e^{-2\pi/h}.
  \label{energy_gain}
\end{equation}
For comparison, the DGR standing wave state has the energy gain $\mu^4
e^{-3\pi/h}$.  The factor of $e^{-\pi/h}$ difference is explained by the
fact that DGR state involves only two patches on the Fermi surface with a
relative area of $e^{-\pi/h}$.

Alternatively, it might be energetically more favorable for the patches on
the Fermi sphere to be overlapping.  In this case, a given particle (or
hole) near the Fermi sphere participates in many pairings simultaneously.
It could be expected that the binding energy of each individual pair is
lower than the value it would have in the non-overlapping case, but
nothing can be said about the total energy of the system.  Indeed, our
preliminary estimation shows that the energy gain is still parametrically
given by Eq.\ (\ref{energy_gain}).  Further investigation is required to
find the true ground state of QCD at very large $\Nc$.

Finally, let us note an interesting possibility that the ground state of
finite-density QCD at very large $\Nc$ might be similar to the
``tomographic Luttinger liquid'' in 2D, advocated by Anderson as the
normal state of high-$T_c$ cuprates \cite{Anderson}.  Such similarity
could stem from the singular interaction between fermions moving in the
same directions, which is also characteristic of tomographic Luttinger
liquids.  As in the case of the latter, one could expect the chiral
symmetry to be unbroken, but the chiral response to be singular at
wavenumber $2\mu$.

\acknowledgements

The authors thank M.~Alford and K.~Rajagopal for stimulating
discussions.  DTS thanks 
P.A.~Lee for helpful conversations.  This work is supported in part by
funds provided by the U.S.\ Department of Energy (DOE) under cooperative
research agreement \#DE-FC02-94ER40818.  The work of ES is also supported
in part by funds provided by the National Science Foundation (NSF) through
the NSF Graduate Fellowship.

\end{document}